\documentclass[%
 reprint,
 amsmath,amssymb,
 aps,
]{revtex4-2}

\usepackage{graphicx}
\usepackage{dcolumn}
\usepackage{bm}
\usepackage{amsmath}

\begin{document}

\preprint{APS/123-QED}

\title{Correlating DC SQUID Performance with the Location of Trapped Magnetic Flux Using Scanning SQUID Microscopy}
\author{B. Xu}
\affiliation{%
Associate of the National Institute of Standards and Technology, Boulder, Colorado 80305, USA
}%
\affiliation{ 
Department of Physics, University of Colorado Boulder, Boulder, Colorado 80309, USA
}%
\author{I. W. Haygood}%
\affiliation{%
National Institute of Standards and Technology, Boulder, Colorado 80305, USA
}%
\author{K. Jackman}
\affiliation{%
SUN Magnetics (RF) (Pty) Ltd, Stellenbosch 7600, South Africa
}%
\author{C. J. Fourie}
\affiliation{%
Department of Electrical and Electronic Engineering, Stellenbosch University, Stellenbosch 7600, South Africa
}%
\author{P. F. Hopkins}
\affiliation{%
National Institute of Standards and Technology, Boulder, Colorado 80305, USA
}%
\author{M. L. Schneider}
\affiliation{%
National Institute of Standards and Technology, Boulder, Colorado 80305, USA
}%

\date{\today}

\begin{abstract}
Holes in the ground planes of superconducting circuits, commonly referred to as moats, are known to be an effective means of mitigating the deleterious effects of residual magnetic flux in superconducting circuits. Previous studies have utilized scanning SQUID microscopy (SSM) to image the effectiveness of various moat geometries and the maximum magnetic fields where they no longer prevent vortices in the ground plane but have not directly correlated the location of flux with circuit performance measurements. In this study we employ SSM to image the effectiveness of various moat configurations in trapping unwanted flux near a DC SQUID, and when the moats are not effective, the location of vortices in the ground plane. During the same cooldown, and without disturbing the circuit, we measure the circuit performance to directly correlate this with the vortex location. We show that there are preferential pinning sites for fluxons regardless of moat configuration, demonstrating the utility of the SSM in provide feedback for circuit layout, modeling, and failure analysis.
\end{abstract}

\maketitle

\section{\label{sec:level1}Introduction}

Superconducting electronics utilizing the magnetic flux quantum, $\Phi_0$, as the information carrier can operate at higher frequencies and with lower power consumption compared to existing CMOS architectures.\cite{1,2,3} Initial architectures relied on switching of hysteretic Josephson junctions (JJs).\cite{4,5,6} More recent and energy efficient architectures using direct-current (DC) biased overdamped, non-hysteretic, JJs have been demonstrated, including Rapid-Single-Flux-Quantum (RSFQ)\cite{3}, energy-efficient-Single-Flux-Quantum (ERSFQ/eSFQ)\cite{7,8}, LR-biased SFQ\cite{9}, and low-voltage RSFQ (LV-RSFQ)\cite{10}. Alternating-current (AC) biased designs which either eliminate static power dissipation on-chip such as the Adiabatic Quantum Flux Parametron (AQFP)\cite{11}, or move the dissipation to room temperature electronics like Reciprocal Quantum Logic (RQL)\cite{12,13,14}. While many different architectures have been demonstrated all of them are susceptible to deleterious effects caused by trapping of fluxons near sensitive circuit components. Perpetual currents induced in nearby superconducting loops disturb operating points, resulting in reduced electrical margins or non-functional circuits.

The density of fluxons, and hence the probability of trapping flux near a circuit element, can be reduced by employing magnetic shielding. However even with significant attenuation achievable only from several layers of ideal aspect ratio shielding (\textgreater $10^{5}$) there will still be dozens of flux quanta in a 1 $\text{cm}^{2}$ area.\cite{15} With modern junction density exceeding $10^{6}$ $\text{cm}^{-2}$ even a relatively low number of fluxons can affect performance.\cite{16} Additionally, flux can be introduced when hooking up electronics or changing measurements meaning that flux mitigation techniques other than shielding are required.

Several active techniques have been proposed to mitigate trapped flux by moving fluxons using thermal gradients\cite{17,18}, using external AC magnetic fields\cite{19}, using pulsed currents to apply Lorentz force to vortices at elevated temperatures\cite{47},  or by varying the thickness of the superconducting film and applying a DC current to exploit the "ratchet effect."\cite{20} A more ubiquitous technique involves patterning holes in the ground plane of circuits, commonly referred to as moats, which act as low-energy sites to preferentially attract flux to a desired area away from critical circuit elements.\cite{21,22,23,24,25,26,27,28} Typically these moats are realized as arrays of rectangles or squares repeated in the ground planes with additional ground plane holes placed surrounding critical circuit elements. Recent studies have shown that perhaps an even more effective approach is to place high aspect ratio (\textgreater 30) rectangular moats near sensitive parts of a layout or in between repeated structures.\cite{29, 39}

The effectiveness of different moat geometries and layouts has been experimentally investigated using various imaging techniques such as nitrogen-vacancy (NV) center imaging\cite{29,30}, magnetic force microscopy (MFM)\cite{31}, and scanning SQUID microscopy (SSM) imaging\cite{32,33}, as well as electrical characterization using shift registers\cite{34,35,36,37}. In this paper we present a study combining both SSM imaging and electrical measurements of DC SQUIDs in a single instrument and single cooldown. This capability allows us to directly correlate the location of fluxons, trapped in moats, or threading the ground plane with changes observed in the critical current, $I_c$, of the SQUID under test (SUT). By enabling a direct measurement of actual flux location on measured circuit performance we demonstrate a valuable tool for validating and providing feedback to simulation models. Additionally, our measurements of four different DC SQUIDs indicate that the physical circuit layout and not just defects in the superconducting film, lead to preferential pinning sites for flux outside of moats, establishing the potential of using SSM as a tool for feedback from failure analysis to best practices for circuit design.

\section{Experiment}

\begin{figure}
\includegraphics[width=\columnwidth]{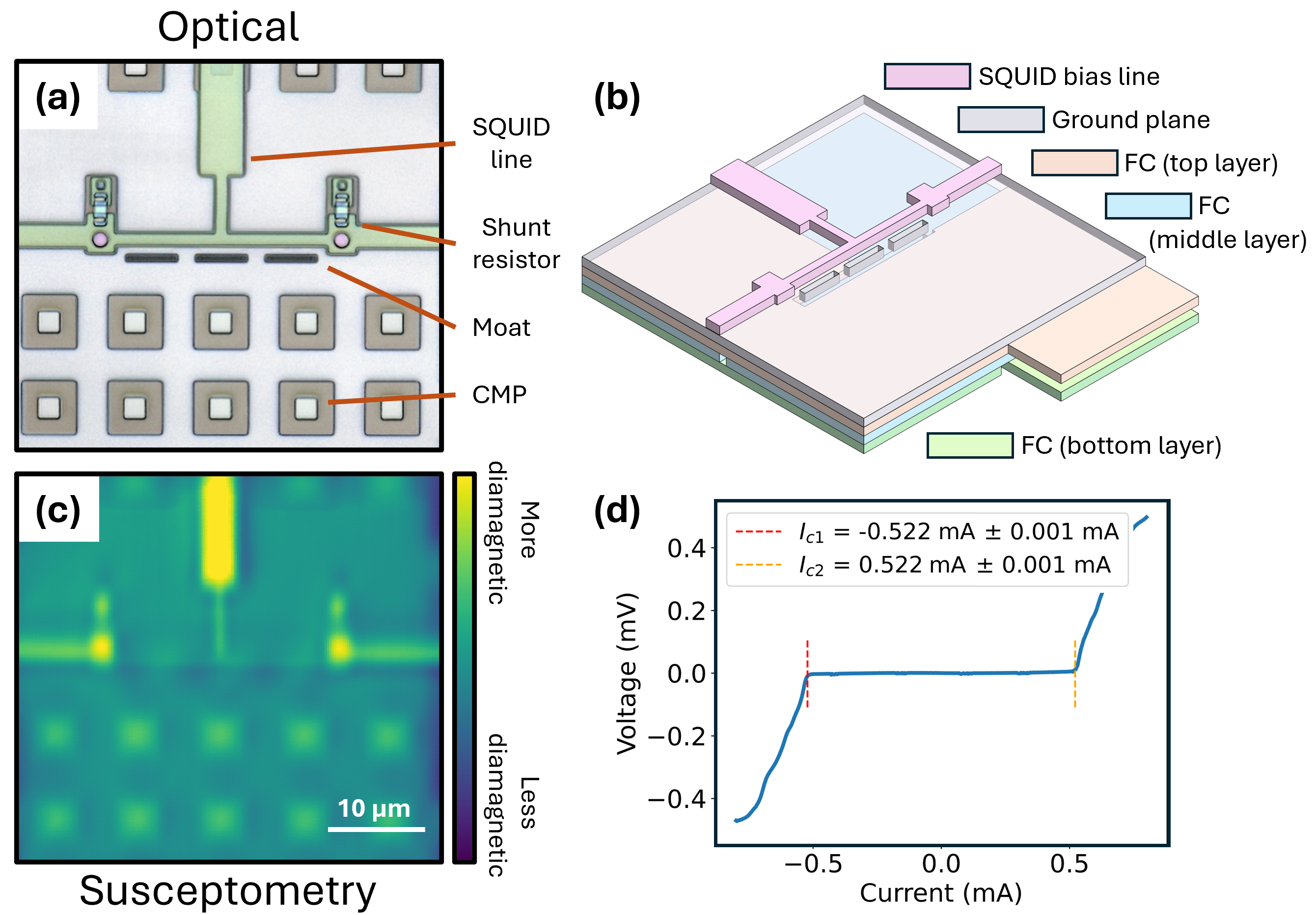}
\caption{\textbf{Device structure and characterization of the DC SQUID with flux-trapping moats.} (a) Optical image of the measured DC SQUID device, with the SQUID bias line, shunt resistors, moats, and CMP fill structures indicated. (b) Illustration of the multilayer device geometry showing the DC SQUID above the ground plane moats above the two-turn field coil (FC). (c) Susceptometry image of the same device region, revealing diamagnetic superconducting elements. (d) Representative current–voltage characteristic of the device cooled in no applied field, with  $I_c$ = 0.522 mA $\pm$ 0.001 mA.}
\label{fig1} 
\end{figure}

To correlate the location of trapped flux or pinned vortices with the electrical characteristics of a superconducting circuit, we measured an integrated circuit that has various moat configurations, i.e. number of moats, aspect ratio, and vicinity, near a number of DC SQUIDs\cite{25,46}. Each SQUID has a designed zero-field critical current $I_c=522~\mu$\text{A} and a loop inductance $L=9.1$~\text{pH} at 4.2 K. It is well known that $I_c$ for a DC SQUID is modulated by applied flux, that the modulation is periodic with $\Phi_{0}$, and that the modulation depth is a function of $I_c$ and $L$. The SQUIDs were designed as flux sensors, with a modulation depth of 27 \% (from 522~$\mu$\text{A} to 379~$\mu$\text{A}).  An optical image of one of these center-tapped DC SQUIDs with three moats is shown in Fig.~\ref{fig1}(a), where an on-chip coil is patterned directly below each SQUID, underneath the main ground plane. The 8-layer MIT Lincoln Laboratory SFQ5ee process\cite{48} was used for fabrication. The cryostat that the measurements were taken in is magnetically shielded and has a residual background flux density of less than 1~$\mu$T, or less than one fluxon per 45~$\mu$m~$\times$~45~$\mu$m area, which almost always results in no fluxons in the moats or ground plane near the SQUID under test (SUT). To add a controlled number of trapped fluxons, the SQUID under test was cooled in a small applied magnetic field.  During this process the temperature of the chip was raised above the critical temperature,  $T_c$~=~9.2~K, of niobium, a current was then applied to either an on chip, two-turn field coil (FC) located beneath the SQUID and moats, depicted in Fig.~\ref{fig1}(b), or through a copper field coil located beneath the PCB to which the chip was wire bonded. The chip was then cooled at a rate of 5~K/min to a base temperature of 4.2~K. The field coil current was then turned off for all subsequent measurements. 

\section{Results}

\begin{figure*}
\includegraphics[width=1.0\textwidth]{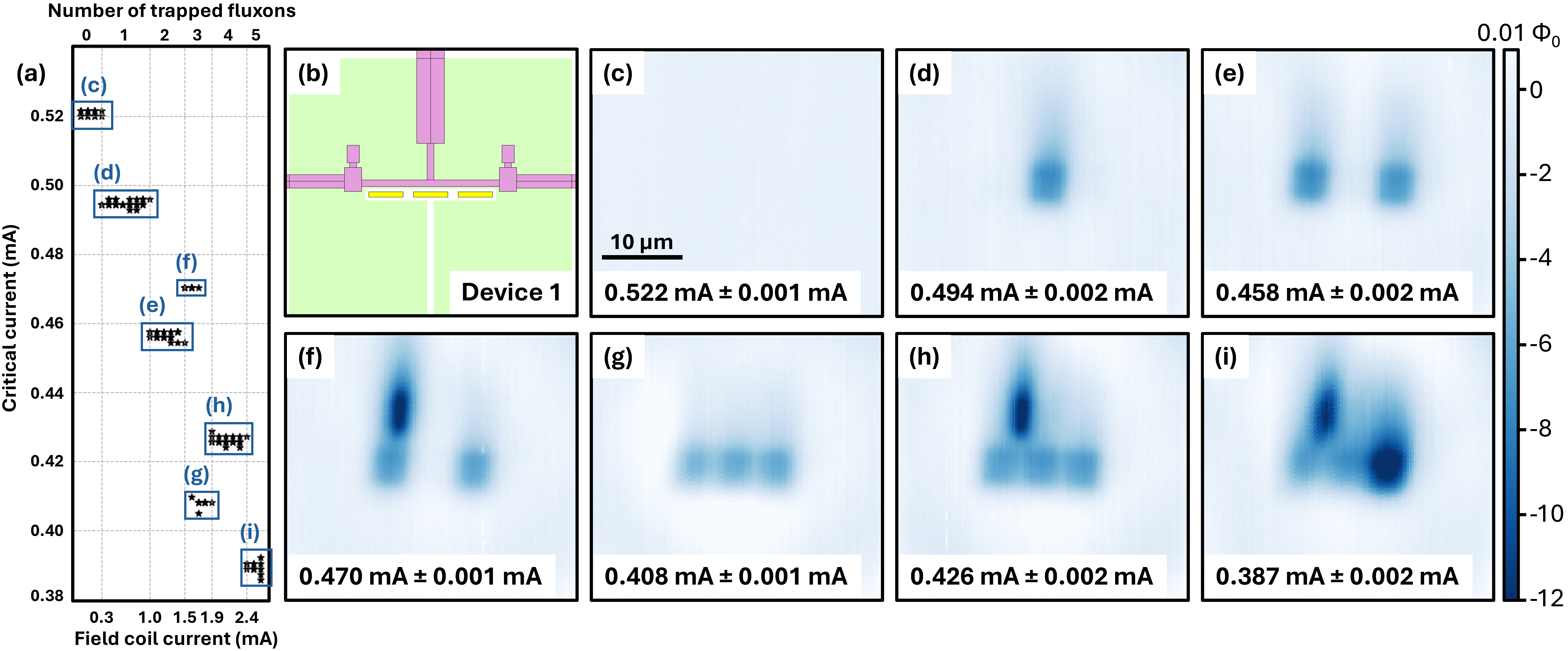}
\caption{\textbf{Correlation between trapped flux and DC SQUID critical current.} (a) Critical current of the three-moat device as a function of the applied field coil current. The top axis indicates the number of fluxons trapped in the scanned region. (b) Corresponding device schematic, with pink SQUID bias lines, yellow moats, and the lower field-coil layer in green.  (c)-(i) Scanning SQUID magnetometry images of the representative configurations corresponding to the labeled boxes in (a). The dark localized features indicate trapped fluxons. The number of trapped vortices increases as the SUT is cooled in higher applied field. Vortices trapped outside the moats are observed in (f), (h) and (i). The measured $I_c$ for each configuration is listed in the bottom left of each image.
}
\label{fig2}
\end{figure*}

\begin{figure}
\includegraphics[width=\columnwidth]{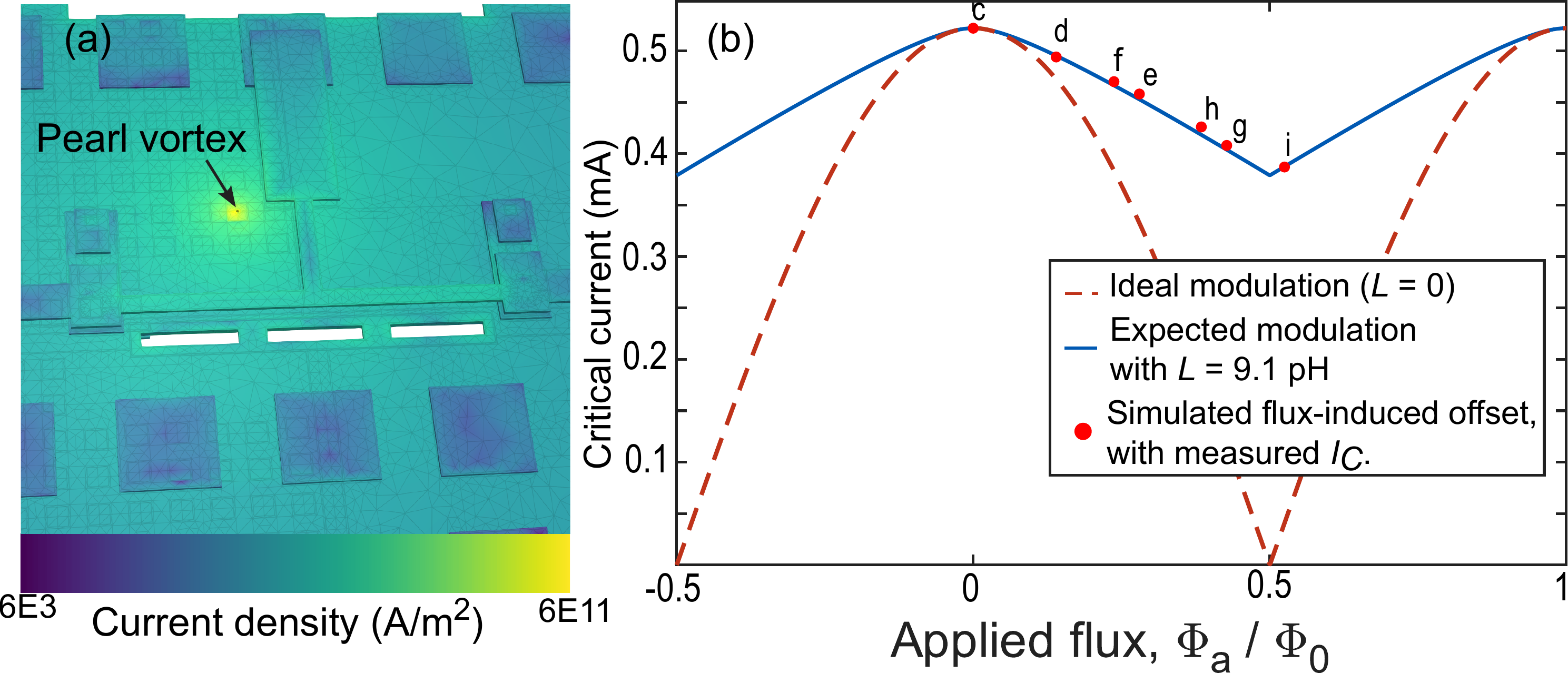}
\caption{\textbf{Simulated effect of moat-trapped fluxons and pinned vortex on DC SQUID response.} (a) Simulated dc current distribution when a Pearl vortex is pinned as measured in SSM imaging for Device 1. (b) Critical current modulation of DC SQUID as a function of applied flux for Device 1, with flux offsets for each moat and the pinned vortex applied to match the combinations shown in Fig.~\ref{fig2}.}
\label{fig3} 
\end{figure}

Immediately after base temperature was reached, a current-voltage (IV) curve was measured for the SUT . A representative IV curve with a critical current of 0.522 mA ± 0.001 mA is shown in Fig.~\ref{fig1}(d) . The negative, $I_{c1}$,  and positive, $I_{c2}$, critical currents were extracted by numerically differentiating the IV curve and identifying where the slope rapidly changes from close to zero to a finite value. This method of extracting the critical current of the SQUID was used because parasitic capacitance and inductance, due to the chip layout,  result in IV curves which deviate from a resistively shunted junction model often used to fit overdamped devices.  The critical current of the SUT was then defined as $I_c=(\lvert I_{c1} \rvert + \lvert I_{c2} \rvert)/2$. The uncertainty in the extracted $I_c$ was estimated using a parametric bootstrap method\cite{45}. Immediately after the electrical measurement, both magnetometry and susceptometry SSM images were taken simultaneously. The SSM sensor used consists of a square pickup loop with inner and outer widths of 0.5 $\mu$\text{m} and 1.5 $\mu$\text{m}, integrated with a concentric circular susceptometry coil having inner and outer diameters of 3 $\mu$\text{m} and 6 $\mu$\text{m}, respectively. The magnetometry channel records the DC flux coupled to the pickup loop, while the susceptometry channel measures the ac flux response generated by an excitation current applied to the susceptometry coil\cite{38, 40, 41}. This process was repeated five times at each field coil current to account for the stochastic nature of flux trapping.

 A three dimensional illustration of the device is shown in Fig.~\ref{fig1}(b), where circuit elements, including the chemical mechanical planarization (CMP) fill structures, shunt resistors, Josephson junctions, and vias connecting the field coil layers are not shown in the illustration for clarity. The contrast in the susceptometry image, shown in Fig.~\ref{fig1}(c), reflects the diamagnetic response of the superconducting elements on the chip, as well as the height differences among different device layers. The regions with the weakest diamagnetic response correspond to the ground plane, which is made of niobium and is located on top of the on-chip field coil. The strongest diamagnetic response is observed near the DC SQUID lines and the shunt-contact regions, which are patterned on the top layer of the chip. The intermediate contrast corresponds to the CMP fill structures, which were added to improve surface uniformity and planarize the device during fabrication. The darkest slits near the SQUID bias line correspond to the moats. These features are less pronounced because the moat width is comparable to the spatial resolution of our sensor. Overall, the susceptometry scan resolves the multilayer superconducting geometry of the device and is in good agreement with the optical image, allowing us to precisely correlate coordinates from the magnetometry data with the actual on-chip location.

 For the following discussion we detail measurements of the DC SQUID with 3 moats described in Fig.~\ref{fig1}. The on-chip field coil was designed to produce a total flux equal to $\Phi_{0}$ through ground plane when coil current is approximately 0.5~mA. A top down representation of the SUT is shown in Fig \ref{fig2}(b) and acts as an aid to the eye for the subsequent discussion and sub-figures. We can verify the absence of any trapped flux through electrical measurements of $I_{c}$, shown in Fig.~\ref{fig2}(a) where there is no measurable change in the critical current of the SQUID, and by magnetometry imaging as demonstrated in Fig.~\ref{fig2}(c) where a uniform low field background is evident. Note that the labeled flux images in Fig.~\ref{fig2}(c)-(i) correspond to the measured critical currents labeled with the same character in Fig.~\ref{fig2}(a). 

As the field coil current is increased above 300 $\mu$A, the applied flux is sufficient to trap one fluxon within the scanned region, with the fluxon occupying one of the moats near the DC SQUID line. Figure~\ref{fig2}(d) shows the case in which a single fluxon is trapped in the center moat. The trapped fluxon appears elongated along the vertical direction in the SSM scan due to the point-spread function (PSF) associated with the sensor geometry \cite{42, 43}. When one fluxon is trapped in a moat, the critical current is consistently reduced by approximately 6\% from its maximum value (Fig.~\ref{fig2}(a) box d). The degradation of the critical current arises because magnetic flux from a fluxon trapped in the moat couples into the SQUID loop, changing the relative phase across the two Josephson junctions and thereby modifying the interference between their supercurrents\cite{44}, and the result matches that obtained through a field simulation of the full circuit\cite{25}. For field-coil currents between 1.0 mA and 1.5 mA, two fluxons are trapped within the scanned region. As shown in Fig.~\ref{fig2}(e), the two fluxons occupy the two outer moats with the center moat empty. This configuration represents the lowest energy configuration for two trapped fluxons due to the repulsive force between them. In this configuration, the critical current is further reduced relative to the single-fluxon case due to additional external flux threading the SQUID.

At field-coil currents above 1.5 mA, two distinct flux configurations corresponding to two different measured values of $I_c$ are observed. The higher measured value of $I_c$ corresponds to the flux configuration imaged in Fig.~\ref{fig2}(f) where the two outer moats are occupied with a single fluxon each and there is a fluxon not contained by a moat, existing as a Pearl vortex on the other side of the SQUID loop. The flux from the Pearl vortex and the moats are pointing out of the ground plane, but because the Pearl vortex is on the other side of the SQUID, the flux threading the SQUID opposes the flux from the fluxons trapped in the moats, resulting in less external flux in the SQUID and  larger measured $I_c$. The lower values of $I_c$ observed at the same field coil current occur when all three moats contain a trapped flux quantum and no Pearl vortex is present as shown in Fig.~\ref{fig2}(g). In this state there is no opposing flux from a vortex on the upper side of the SQUID resulting in more external flux and larger suppression of $I_c$.

At larger field-coil currents, the device enters the four- and five-fluxon regimes. In Fig.~\ref{fig2}(h), all three moats remain occupied, while an additional fluxon is trapped outside the moat array. This off-moat trapping site is very close to the off-moat fluxon locations observed in Fig.~\ref{fig2}(f) and results in an increase in $I_c$ for the same reason as discussed previously. Additionally, a similar vortex location indicates a reproducible local pinning site near the device. In Fig.~\ref{fig2}(i), a fifth fluxon is trapped, with two fluxons consistently occupying the rightmost moat. This configuration is likely favored by repulsion from the off-moat fluxon, which drives additional fluxons toward the moat farthest from it. 

A full three-dimensional calculation of the inductance of each moat and the Pearl vortex, as well as the coupling from each moat and the Pearl vortex to the DC SQUID loop inductance, was performed with InductEx\footnote{This does not imply endorsement by NIST or that the product is the best available for the purpose.}. From these results, the flux coupled to the SQUID loop is calculated for each flux trapping position. The critical current for each trapped flux combination in Fig.~\ref{fig2} is then calculated from the combined flux coupling as shown in Fig.~\ref{fig3}. From simulation, the flux coupled to the SQUID loop inductor is 0.14$\Phi_0$ per fluxon in the left and right moats, 0.147$\Phi_0$ for the center moat, and -0.043$\Phi_0$ for the Pearl vortex (it induces flux in the opposite direction as the moats). The simulated and measured results agree very well, which confirms that pinned vortex does indeed have a significant influence on the SQUID.

\begin{figure}
\includegraphics[width=1.0\columnwidth]{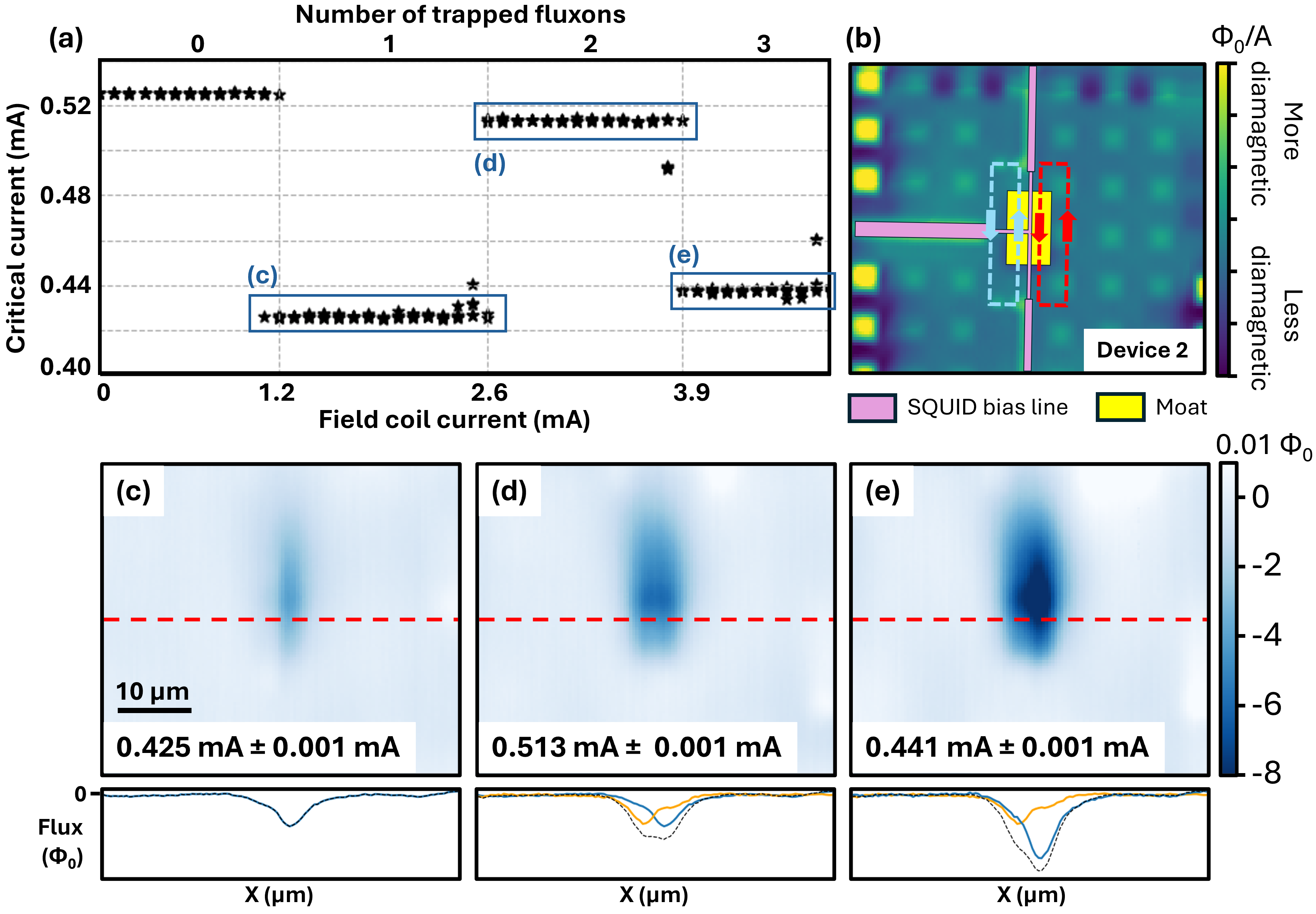}
\caption{\textbf{Spatially distinct flux trapping within a single moat.} (a) Critical current $I_c$ of the one-moat device measured after temperature cycling, plotted as a function of the applied field current and corresponding number of fluxons trapped in the scanned region. (b) Scanning SQUID susceptometry image overlaid with a schematic of the device layout. A single rectangular moat, shown in yellow, is positioned beneath the SQUID bias line, shown in pink. The blue and red loops illustrate the local screening-current paths associated with vortices trapped in the two sections of the moat. (c)-(e) Scanning SQUID magnetometry images of the same device after temperature cycling in different applied fields, showing one, two, and three trapped fluxons. The critical current measured for each trapped-flux configuration is labeled in the corresponding image. The red dashed line indicates the line cut used to obtain the profiles plotted below the magnetometry images. Black dashed curves show the total line-cut signal, while the blue and orange curves separate the contributions from fluxons on the left and right sides by subtracting the preceding trapped-flux configuration.
}
\label{fig4} 
\end{figure}

The results in Fig.~\ref{fig2} show that the effect of trapped flux on the DC SQUID strongly depends on the location of the fluxons relative to the circuit. This raises a broader question of how nearby circuit elements modify the trapping behavior of an engineered moat. To investigate this, we examined a device that contains a single rectangular moat positioned directly beneath the SQUID bias line. Although the moat is continuous in the ground plane, the bias line crosses above it on a different superconducting layer, providing a test of whether multilayer circuit geometry can divide a nominally continuous moat into distinct flux-trapping regions. Figure \ref{fig4}(a)  is a plot of the SQUID response as a function of field coil current. The oscillatory response of the critical current as a function of the field can be explained by looking at the specific current paths that current can flow in this device, as theorized in Ref. 25. The two different current paths are shown in Fig. \ref{fig4}(b) as blue and red dashed paths, overlaid on a susceptometry image with the device bias lines and moats filled in for clarity. The single rectangular moat (yellow box) can be seen directly beneath a DC SQUID similar to the device studied in Fig. \ref{fig2}. Note that the SUT was rotated by 90 degrees to improve the spatial resolution along the horizontal direction and compensate for the stretching due to the PSF mentioned earlier. 

Fig. \ref{fig4}(c)–(e) show magnetometry images of the same region as Fig. \ref{fig4}(b) after temperature cycling in increasing applied fields, along with line profiles of the DC flux across the center of the moat (red dashed line), corresponding to configurations with one, two, and three trapped fluxons. For the one-fluxon configuration shown in Fig. \ref{fig4}(c), the fluxon is localized on the right side of the moat, and the critical current is reduced by 19 percent. The screening current associated with the trapped fluxon follows a path around the right section of the moat in the superconducting ground plane, transfers to the upper superconducting layer through the Josephson junction, and then crosses the SQUID bias line before reaching the opposite end of the moat. The line cut also shows a clear, single peak in the flux, confirming the localization on the right side of the SQUID in Fig. \ref{fig4}(c).

At higher fields, when a second fluxon is trapped, as shown in Fig. \ref{fig4}(d), the critical current almost completely recovers to within 2 percent of the zero field case. This is consistent with fluxons located on opposite sides of the SQUID bias line, each with the same polarity. This is further confirmed by examining the line cut where two distinct peaks are observed. Here the orange curve (left fluxon) was obtained by subtracting the curve from Fig. \ref{fig4}(c) from the total flux response (black dashed curve). The slight depression in  $I_c$ is due to slight asymmetry between the two current paths due to the layout or lithographic imperfections.  The corresponding screening-current paths are illustrated by the red and blue dashed lines in Fig. \ref{fig4}(b). These results confirm that the SQUID bias line acts as a bridge for the screening current and practically separates the single rectangular moat into two regions that can trap fluxons independently. 

Finally, in Fig. \ref{fig4}(e), two fluxons are trapped in the right section of the moat while one fluxon is trapped in the left section. In this configuration $I_c$ is once again suppressed, but slightly less than the one fluxon configuration, due to asymmetries in the coupling between the left and right current paths.  The evidence of the fluxons configuration is also evident in the line cut where two distinct peaks are observed with the right peak approximately two times the magnitude of the left-side signal. A similar method as above was used to resolve the two peaks where the orange curve from Fig. \ref{fig4}(d)  was subtracted from the total flux response to generate the blue curve. These measurements demonstrate that the trapping behavior of a moat is determined not only by its geometry in the ground plane but also by nearby circuit elements in other superconducting layers. In this case, the SQUID bias line provides a bridge for the screening current and effectively divides a single rectangular moat into two independently occupied trapping regions, consistent with the behavior predicted in Ref. 25.

\begin{figure}[b]
\includegraphics[width=\columnwidth]{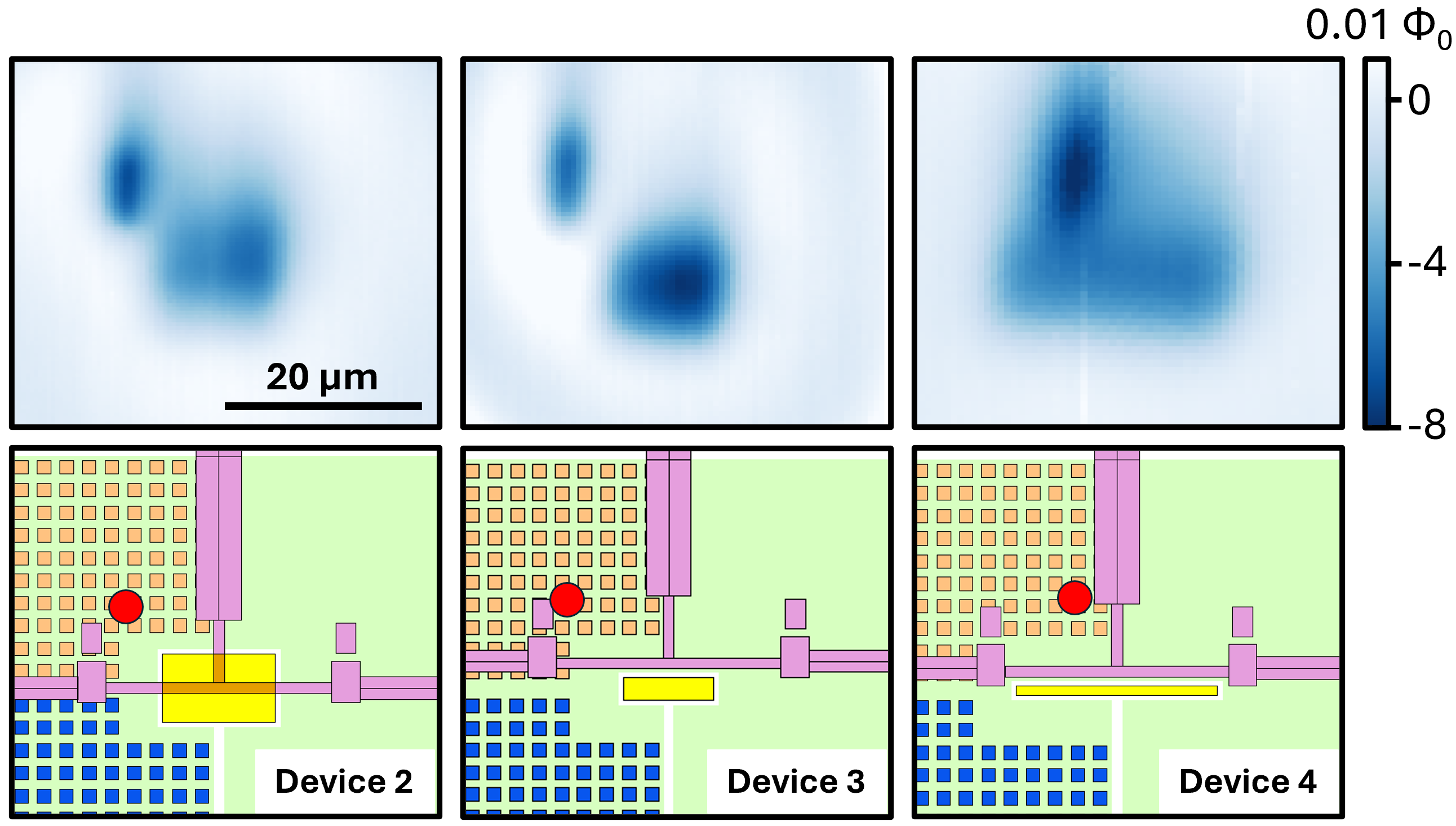}
\caption{\textbf{Preferred trapping locations of the first off-moat Pearl vortex under local field cooling.} Upper row: scanning SQUID magnetometry images showing the first off-moat Pearl vortex in three devices with different moat geometries, with the applied field generated by the on-chip field coil. Lower row: Corresponding device schematics, with pink: SQUID lines, yellow: moats, green: lower field-coil layer, orange: vias connecting the upper and middle field-coil layers, dark blue: vias connecting the middle and bottom field-coil layers. The red dots mark the observed Pearl vortex positions. The first off-moat vortex is consistently trapped near the upper via-array region.
}
\label{fig5} 
\end{figure}

\begin{figure*}
\includegraphics[width=0.9\textwidth]{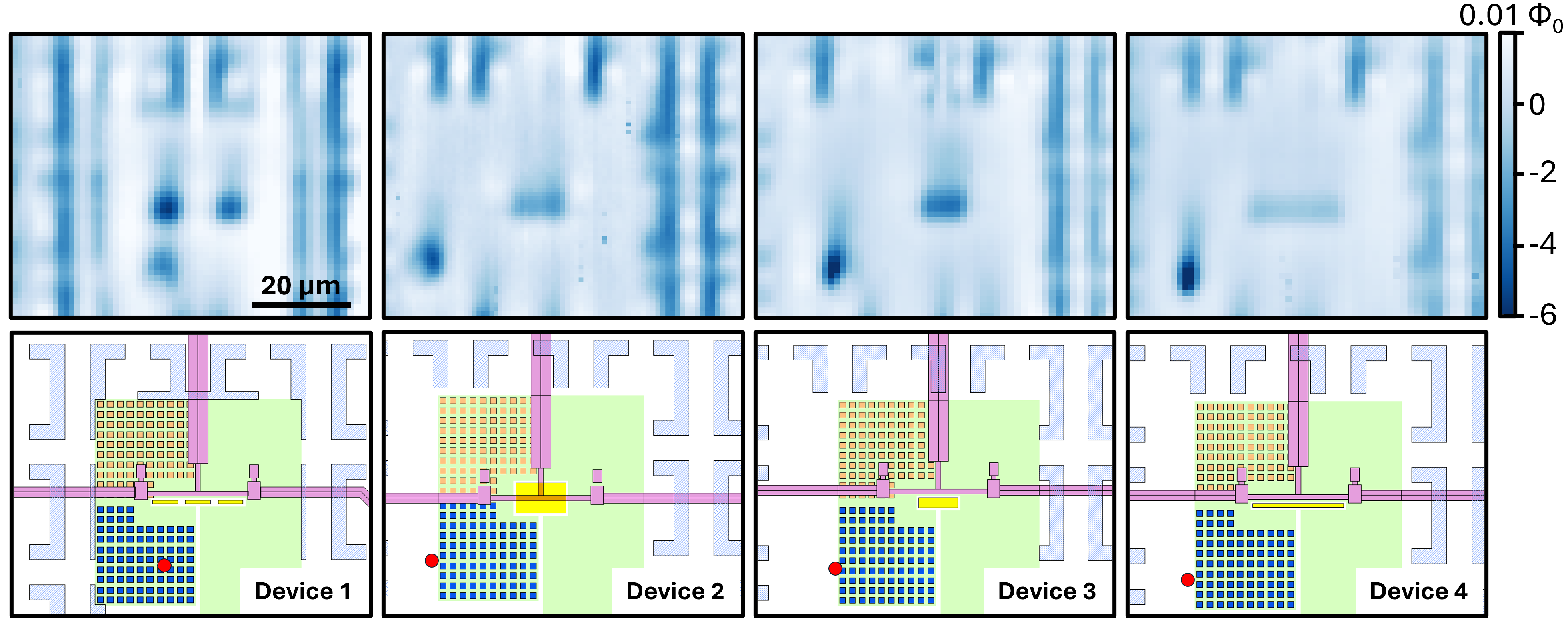}
\caption{\textbf{Preferred trapping locations of the first off-moat Pearl vortex under global field cooling.} Upper row: scanning SQUID magnetometry images showing the first off-moat Pearl vortex in four devices with different moat geometries, with the applied field generated by a global coil beneath the chip. The field applied in each image, from left to right, is 13.4 $\mu$T, 6.8 $\mu$T, 6.8 $\mu$T, 6.8 $\mu$T. Lower row: corresponding device layouts, using the same color scheme as in Fig. \ref{fig5}, plus light blue indicating the outer moats surrounding the device area. The first off-moat vortex is consistently trapped near a field-coil via array or edge, indicating preferred pinning associated with the multilayer device structure.
}
\label{fig6} 
\end{figure*}

Looking again at the three moat device of Fig.~\ref{fig2}, another notable feature observed is that the off-moat vortices in Fig.~\ref{fig2}(f) \ref{fig2}(h), and \ref{fig2}(i) appear at nearly the same location, suggesting the presence of a local low-energy trapping site near the device. To examine this finding further, we examined three additional devices with different moat configurations under two field-cooling conditions and identified the location of the first Pearl vortex trapped outside the engineered moats. In Fig.~\ref{fig5}, the applied field is generated by the local on-chip field coils, whereas in Fig.~\ref{fig6} it is generated by the global field coil beneath the chip. The locations of the trapped fluxons are captured by overlapping the magnetometry, susceptometry images with the SUT layouts.

The devices shown in Fig.~\ref{fig5} share the same basic multilayer structure, consisting of a two-turn field coil across three superconducting layers beneath the ground plane and the SQUID wiring above it. The field-coil turns are connected between superconducting layers through via arrays, shown as orange and blue squares, located to the left of the active device region. Despite the different moat geometries, the first off-moat vortex in all three devices appears at nearly the same location in the upper via-array region, when the applied field is generated by the on-chip field coil. This reproducibility indicates that, at fields where moats do not prevent off-moat trapping, the preferred trapping location is governed primarily by the physical device layout rather than by randomly distributed defects in the niobium ground plane. 

Because the field generated by the local on-chip coil may not represent the uniform magnetic environment typically experienced by superconducting circuits, we repeated the measurements using a global field coil beneath the chip. Results are shown in Fig.~\ref{fig6}. In this configuration, no current flows through the local on-chip field coil, and the applied field is approximately uniform over the dimensions of the device. The applied field was increased until the first off-moat vortex was observed; from left to right, the corresponding applied fields were 13.4~$\mu$T, 6.8~$\mu$T, 6.8~$\mu$T, and 6.8~$\mu$T, respectively. The preferred vortex locations shift away from the upper via-array region and instead occur near the lower via-array region or other edges of the multilayer on-chip field coil. The disappearance of the upper-left-site preference indicates that the tightly focused field of the local on-chip field coil influences the selection of the preferred trapping site. One possible interpretation is that current crowding and changes in the current direction near the interlayer connections distort the local magnetic-field distribution and lower the vortex-entry barrier in selected regions. Because the upper-left via array is located closer to the ground plane layer, the resulting field distortion may couple more strongly to the ground plane and thereby create a preferred pinning site in this region across all tested devices. At the same time, the continued localization of vortices near via arrays and layer boundaries under global field cooling demonstrates that the multilayer structure itself produces a nonuniform pinning landscape.

While the Figs.~\ref{fig5} and Figs.~\ref{fig6} establish a clear correlation between preferred trapping sites and the multilayer device geometry, rather than with the moat geometry alone, the microscopic origin of these preferred sites remains unclear, and several mechanisms may contribute. The via arrays and layer transition can introduce local topography, film-thickness variation, or stress, leading to spatial variation in the superconducting properties that can modify the local vortex-pinning energy. In addition, thermal gradient during cooldown, together with variations in the superconducting properties and critical temperatures among layers may cause some layers to become superconducting before others. The geometry of the layers that have already transitioned may dominate the available paths for flux expulsion and trapping, thereby modifying the eventual vortex locations. Determining the relative importance of these effects and the microscopic origin of the preferred trapping sites will require further investigation.

\section{Conclusion}
We have demonstrated the utility of using SSM to investigate flux trapping in superconducting moats and in the ground plane near a superconducting circuit primitive, and to validate the three-dimensional field-solver calculation of the coupling of fluxons trapped in moats and as Pearl vortices to superconducting circuits. The results show precisely how and where moats are filled and, when they are no longer effective, the location of Pearl vortices in the ground plane. We further show that the trapping behavior of engineered moats is influenced by nearby circuit elements in other superconducting layers, and that vias, layer transitions, edges, and current-induced magnetic fields can create additional preferred trapping sites outside the moats. These results demonstrate that effective flux management requires consideration of the complete multilayer circuit geometry and current distribution, rather than moat design alone. More broadly, SSM provides a powerful tool for validating flux-trapping models, identifying unintended pinning sites, and providing feedback for the design and failure analysis of superconducting circuits. Future work will extend this approach to more complex superconducting circuit elements, including DC-to-SFQ converters, Josephson transmission lines, flux shuttles, and neuromorphic primitives.

\bibliography{Main_aps}

\end{document}